# Dynamical Model of Binary Asteroid Systems Using Binary Octahedrons

Yu Jiang[1, 2], Hexi Baoyin[1], Mo Yang[1]

1. School of Aerospace Engineering, Tsinghua University, Beijing 100084, China

2. State Key Laboratory of Astronautic Dynamics, Xi'an Satellite Control Center, Xi'an 710043, China

corresponding author:

Y. Jiang (✉) e-mail: jiangyu_xian_china@163.com

H. Baoyin(✉) e-mail: baoyin@tsinghua.edu.cn

**Abstract**. We used binary octahedrons to investigate the dynamical behaviors of binary asteroid systems. The mutual potential of the binary polyhedron method is derived from the fourth order to the sixth order. The irregular shapes, relative orbits, attitude angles, as well as the angular velocities of the binary asteroid system are included in the model. We investigated the relative trajectory of the secondary relative to the primary, the total angular momentum and total energy of the system, the three-axis attitude angular velocity of the binary system, as well as the angular momentum of the two components. The relative errors of total angular momentum and total energy indicate the calculation has a high precision. It is found that the influence of the orbital and attitude motion of the primary from the gravitational force of the secondary is obvious. This study is useful for understanding the complicated dynamical behaviors of the binary asteroid systems discovered in our Solar system.

## 1 Introductions

The discovery of the binary asteroid systems in our solar system brings a great interest to the dynamics around irregular celestial bodies. Several previous literature has studied the dynamics of the binary systems. Vasilkova (2005) used a triaxial ellipsoid to model the irregular shape of asteroid, and calculated the dynamical behavior of a massless particle orbiting around the relative equilibrium of the triaxial ellipsoid. Lindner et al. (2010) used the massive line segment and a mass particle to model the primary and the secondary of the binary asteroid system, respectively. They discussed the synchronous orbit, chaotic orbits, unstable periodic orbits, and spin-orbit coupling of the system. Liu et al. (2011) investigated the equilibria and orbits for a massless particle in the potential of a rotating cube. Najid et al. (2011) used a straight elongated inhomogeneous segment to model the gravitational potential of the asteroid, and calculated the trajectories and Poincaré sections of a massless particle. Liu et al. (2013) investigated the surface motion of a massless particle on a rotating cube. Jiang et al. (2015) investigated the variety of the relative equilibria in the potential of an asteroid, which can help to understand the dynamical behaviors of the large-size-ratio synchronous binary systems. Ferrari et al. (2016) used three particles to model the binary asteroids, two of them are used to model the gravity of the primary, and the other one is used to model the gravity of the secondary. They refer to this model as the patched three-body problems, and calculated the equilibrium points and orbits around the equilibrium points for this model. These studies can help us to understand the dynamical behaviors of a tiny moonlet orbiting around an asteroid.

However, the size and mass of the moonlet in the binary asteroid systems are not zero, the assumption of the massless particle of the moonlet will cause to the loss of some important dynamical characteristic of the binary systems, including the spin-orbit locked of the moonlet, the topological cases of the relative equilibria, the motion stability of the moonlets, etc. Besides, the shapes of the bodies in the binary asteroid system are irregular. The assumption of the particle masses or sphere cannot model the attitude motion of the systems. Thus the spin-orbit locked, the escape of the moonlet or not, and the resonance of the binary asteroid systems cannot be investigated. Some literature used more suitable model to investigate the dynamics of the binary systems. The most common model is used by the two finite straight segments to simulate the gravity of the binary systems. Guirao et al. (2011) used the two finite orthogonal straight segments to approximate the irregular binary systems, and studied the nonlinear stability of the equilibrium points of the system. Jain and Sinha (2014a) studied the positions and stability of the equilibrium points in the restricted three-body problem with the assumption of both finite straight segments. Under the same assumption, Jain and Sinha (2014b) investigated the non-linear stability of L4 for a massless particle. Blaikie et al. (2014) also used two massive line segments to model the binary systems; they derived the potential energy, kinetic, force, and torque of the system, and presented the unstable periodic orbits and chaotic orbits of the systems. Shang et al. (2015) used the double ellipsoids to model the gravitational environment of the synchronous binary systems, and calculated the periodic orbits for a massless particle in the systems. Nadoushan and Assadian (2016) studied the spin-orbit resonance of the binary asteroids with assuming the shape model of the two asteroids as a sphere and an ellipsoid. The Poincare section for a fictitious binary asteroid is calculated. Elshaboury et al. (2016) used two triaxial rigid bodies to model the two primaries in the restricted three-body problem, and calculated the positions and stability of the equilibrium points in a special case. They also found three unstable collinear equilibria.

In this paper, we use two octahedrons to model the irregular shapes of binary asteroid systems. With this model, irregular shapes, relative orbits, attitude angles, and the angular velocities of the binary asteroid system can be calculated. The mutual potential, force terms and torque terms of the system are calculated by the method of two homogeneous polyhedral. The dynamical equation is integrated by the 7/8-order Runge-Kutta method. This paper is organized as follows: Section 2 deals with motion of the Full Two-body Problems. Section 3 presented the simulation of the binary asteroid system using the binary octahedron. Section 4 gives a brief review of this study.

**2 Motion Equations of the Full Two-Body Problems**

We focus on the binary asteroidal system which comprised of two irregular-shaped bodies. Let $\beta_l (l=1,2)$ be the $l$-th body, $G$ be the gravitational constant, $\mathbf{r}_l$ be the position vector from the origin of the inertial reference frame to the center of mass of the $l$-th body $\beta_l$, $dm(\mathbf{D}_l)$ be the mass element at $\beta_l$, $\mathbf{D}_l$ be the position vector from the center of mass of $\beta_l$ to the mass element, $\rho(\mathbf{D}_l)$ be the density at the mass element which satisfies $dm(\mathbf{D}_l) = \rho(\mathbf{D}_l) dV(\mathbf{D}_l)$, $dV(\mathbf{D}_l)$ be the volume element, $\mathbf{p}_l = m_l \dot{\mathbf{r}}_l$ be the linear momentum vector of $\beta_l$, $\mathbf{q}_l = \mathbf{A}_l \mathbf{D}_l + \mathbf{r}_l$ be the position vector from the origin of the inertial reference frame to the mass element $dm(\mathbf{D}_l)$, $m_l$ means the mass of $\beta_l$, $\mathbf{A}_l$ be the attitude matrix from the inertial reference frame to the principal reference frame of $\beta_l$, $\mathbf{K}_l = \mathbf{r}_l \times \mathbf{p}_l + \mathbf{A}_l \mathbf{I}_l \Phi_l$ be the angular momentum vector of $\beta_l$, $\mathbf{d}_l = \mathbf{A}_l \mathbf{D}_l$, and $\mathbf{G}_l = \mathbf{I}_l \Phi_l$. Using arbitrary vector $\mathbf{v} = \begin{bmatrix} v_x, v_y, v_z \end{bmatrix}^T$, the antisymmetric matrix of $\mathbf{v}$ is denoted as

$$\hat{\mathbf{v}} = \begin{pmatrix} 0 & -v_z & v_y \\ v_z & 0 & -v_x \\ -v_y & v_x & 0 \end{pmatrix}. \tag{1}$$

Denote $\mathbf{R} = \mathbf{A}_2^T \mathbf{r}$, $\mathbf{p} = m\dot{\mathbf{r}}$, $m = \dfrac{m_1 m_2}{m_1 + m_2}$, $\mathbf{P} = \mathbf{A}_2^T \mathbf{p}$, $\mathbf{A} = \mathbf{A}_1^T \mathbf{A}_2$, $\mathbf{\Gamma}_1 = \mathbf{A}^T \mathbf{G}_1$, and $\mathbf{\Gamma}_2 = \mathbf{G}_2$. Then, $\mathbf{G}_k = \mathbf{A}_k^T \mathbf{g}_k = \mathbf{I}_k \Phi_k \ (k = 1, 2)$. And the total mutual gravitational potential energy can be expressed with the parameters in the body-fixed frame of $\beta_2$ as:

$$U = -\int_{\beta_1}\int_{\beta_2} \frac{G\rho(\mathbf{D}_1)\rho(\mathbf{D}_2) dV(\mathbf{D}_1) dV(\mathbf{D}_2)}{\|\mathbf{A}^T \mathbf{D}_1 - \mathbf{D}_2 + \mathbf{R}\|}. \quad (2)$$

The motion equation of the binary asteroidal system (Maciejewski 1995; Naidu and Margot 2014; Jiang et al. 2016, 2017; Wang and Xu 2018) can be written in the body-fixed frame of $\beta_2$ by

$$\begin{cases} \dot{\mathbf{P}} = \mathbf{P} \times \Phi_2 - \dfrac{\partial U}{\partial \mathbf{R}} \\ \dot{\mathbf{R}} = \mathbf{R} \times \Phi_2 + \dfrac{\mathbf{P}}{m} \\ \dot{\mathbf{\Gamma}}_1 = \mathbf{\Gamma}_1 \times \Phi_2 + \boldsymbol{\mu}_1 \\ \dot{\mathbf{\Gamma}}_2 = \mathbf{\Gamma}_2 \times \Phi_2 + \boldsymbol{\mu}_2 \\ \dot{\mathbf{A}} = \mathbf{A}\widehat{\Phi}_2 - \widehat{\Phi}_1 \mathbf{A} \\ \dot{\mathbf{A}}_2 = \mathbf{A}_2 \widehat{\Phi}_2 \end{cases}, \quad (3)$$

where $k = 1, 2$, $\Phi_2 = \mathbf{I}_2^{-1} \mathbf{\Gamma}_2$, and $\Phi_1 = \mathbf{I}_1^{-1} \mathbf{A} \mathbf{\Gamma}_1$.

The kinetic energy of the binary asteroidal system can be written as

$$T = \frac{1}{2}\sum_{k=1}^{2}\left(m_k \|\dot{\mathbf{r}}_k\|^2 + \langle \Phi_k, \mathbf{I}_k \Phi_k \rangle\right), \quad (4)$$

Then, the total energy of this binary asteroidal system can be given by

$$\begin{aligned} H &= T + U \\ &= \frac{1}{2}\sum_{k=1}^{2}\left(m_k \|\dot{\mathbf{r}}_k\|^2 + \langle \Phi_k, \mathbf{I}_k \Phi_k \rangle\right) - \int_{\beta_1}\int_{\beta_2} \frac{G\rho(\mathbf{D}_1)\rho(\mathbf{D}_2) dV(\mathbf{D}_1) dV(\mathbf{D}_2)}{\|\mathbf{A}^T \mathbf{D}_1 - \mathbf{D}_2 + \mathbf{R}\|}. \end{aligned} \quad (5)$$

The gratitational force acting on $\beta_k$ reads

$$\mathbf{f}^k = -G \int_{\beta_1} \int_{\beta_2} \frac{\rho(\mathbf{D}_1)\rho(\mathbf{D}_2) dV(\mathbf{D}_1) dV(\mathbf{D}_2)}{\left\| \mathbf{A}^T \mathbf{D}_1 - \mathbf{D}_2 + \mathbf{R} \right\|^3}. \tag{6}$$

The resultant gravitational torque acting on $\beta_k$ expressed in the inertial space can be written as

$$\mathbf{n}^k = G \int_{\beta_1} \int_{\beta_2} \frac{\left(\mathbf{A}^T \mathbf{D}_1 + \mathbf{R}\right) \times \mathbf{D}_2}{\left\| \mathbf{A}^T \mathbf{D}_1 - \mathbf{D}_2 + \mathbf{R} \right\|^3} \rho(\mathbf{D}_1) \rho(\mathbf{D}_2) dV(\mathbf{D}_1) dV(\mathbf{D}_2). \tag{7}$$

The linear momentum integral and the angular momentum integral of the system are

$$\mathbf{p} = \mathbf{p}_1 + \mathbf{p}_2, \qquad \mathbf{K} = \mathbf{K}_1 + \mathbf{K}_2. \tag{8}$$

Using the infinite series, one can calculate the mutual potential of two homogeneous polyhedral. Werner and Scheeres (2005) presented the series of the mutual potential to the third order as follows:

$$U = G \sum_{a \in \beta_1} \sum_{b \in \beta_2} \rho_a \rho_b T_a T_b \left\{ \left[ \frac{Q}{R} \right] + \left[ -\frac{Q_i w^i}{R^3} \right] + \left[ -\frac{Q_{ij} r^{ij}}{2R^3} + \frac{3Q_{ij} w^i w^j}{2R^5} \right] \right. \\ \left. + \left[ \frac{3Q_{ijk} r^{ij} w^k}{2R^5} - \frac{5Q_{ijk} w^i w^j w^k}{2R^7} \right] + \cdots \right\}. \tag{9}$$

The forth order to the sixth order of the mutual potential can be calculated by

$$U_4 = \left[ -\frac{3Q_{ijkl} r^{ijkl}}{8R^5} + \frac{15Q_{ijkl} r^{ij} w^k w^l}{4R^7} - \frac{35Q_{ijkl} w^i w^j w^k w^l}{8R^9} \right]$$

$$U_5 = \left[ -\frac{15Q_{ijklm} r^{ijkl} w^m}{8R^7} + \frac{70Q_{ijklm} r^{ij} w^k w^l w^m}{8R^9} - \frac{63Q_{ijklm} w^i w^j w^k w^l w^m}{8R^{11}} \right]$$

$$U_6 = \left[ \frac{5Q_{ijklmn} r^{ijklmn}}{16R^7} - \frac{105Q_{ijklmn} r^{ijkl} w^m w^n}{16R^9} + \frac{35Q_{ijklmn} r^{ij} w^k w^l w^m w^n}{16R^{11}} - \frac{231Q_{ijklmn} w^i w^j w^k w^l w^m w^n}{16R^{13}} \right]$$

$$\tag{10}$$

Here $\rho_a$ and $\rho_b$ are densities of the simplexes $a$ and $b$ on the polyhedral, respectively; $T_a$ and $T_b$ are Jacobian determinant of the transformation for the simplexes *a* and *b*, respectively; $Q = \frac{1}{36}$ is the rank-0 tensor of rational numbers, symmetric in all index pairs, and $Q_{i_1 \cdots i_k}$ is the rank-*k* tensor of rational numbers, symmetric in all index pairs; $R = |\boldsymbol{R}|$; $w^i$ is the exporting variable from $\boldsymbol{R}$ and the negative weight of simplexes; $r^{ij}$ and $r^{ijkl}$ are the rank-2 and rank-4 tensors generated by the negative weight of simplex with itself, respectively.

**3 Simulation of the Binary Asteroid Systems Using Binary Octahedrons**

We use two octahedrons to model the irregular shapes of binary asteroidal systems. Generally, the two components of the binary asteroidal system are not the same; we refer to the larger one as the primary (denoted as A in the figures), and the smaller one as the secondary (denoted as B in the figures). The volumes of the octahedrons are set to be $1.8 \times 10^9 \mathrm{m}^3$ and $0.1561 \times 10^9 \mathrm{m}^3$. Densities of them are 2.5 g·cm$^{-3}$. The three axis coordinates of the primary are [-1.0, 1.0]km, [-1.5, 1.5]km, and [-0.9, 0.9]km in x, y, and z axis, respectively. The three axis coordinates of the secondary are [-1.0, 1.0]km, [-0.367879, 0.367879]km, and [-0.318309, 0.318309]km in x, y, and z axis, respectively. To see the shape and gravitation of the octahedrons, we plotted the shape of the primary in Fig. 1, and showed the structure of the gravitational potential for the primary in Fig. 2. The gravitational potential is calculated by the binary polyhedral method (Werner and Scheeres 1997; Tsoulis 2012; D'Urso 2014) to the fourth order. The octahedron in Fig. 2 has the same size and shape with the primary, and its mass is set to be a unit mass because we are only interested in the structure of the gravitational potential. From Fig. 2, one can see that the gravitational potentials in the different coordinate planes are different, and are related to the shape of the octahedron. The maximal value of the gravitational potential is 0.5803 m$^2$s$^{-2}$.

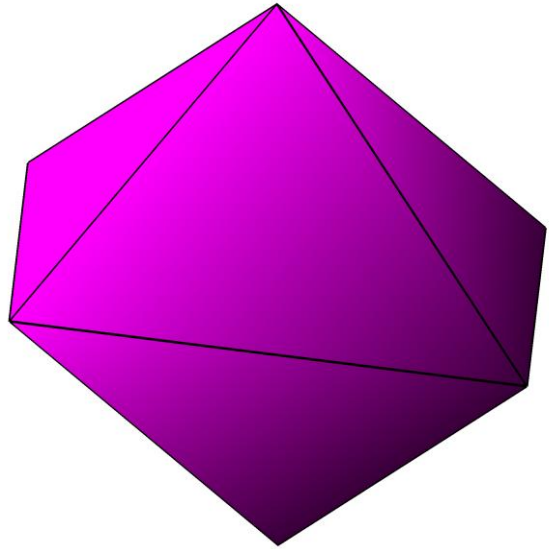

Figure 1. The 3D shape of the primary.

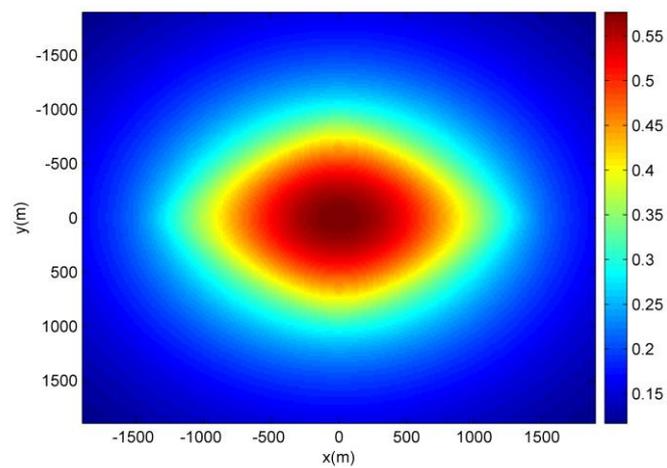

(a)

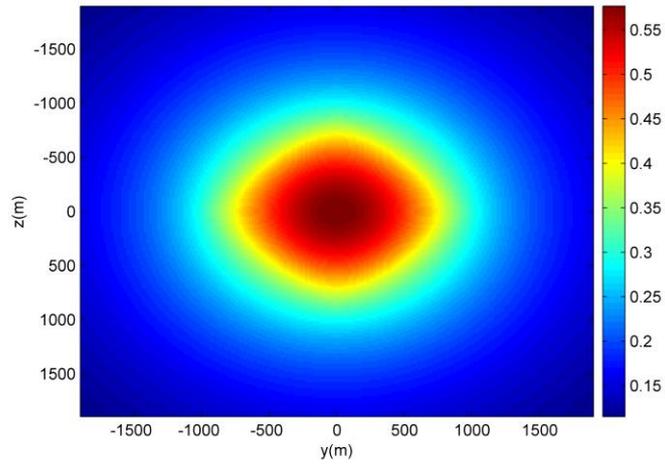

(b)

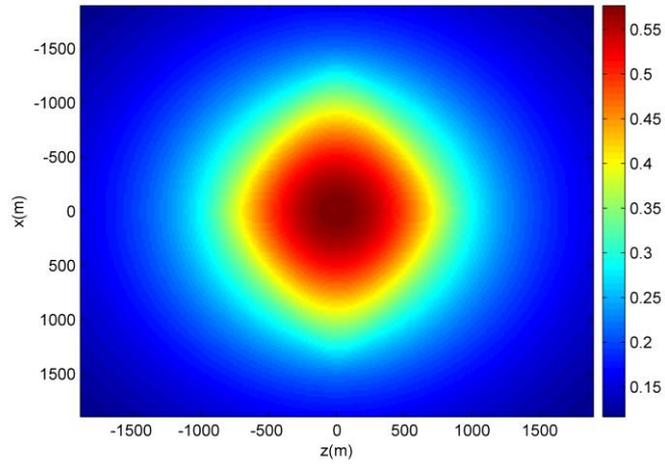

(c)

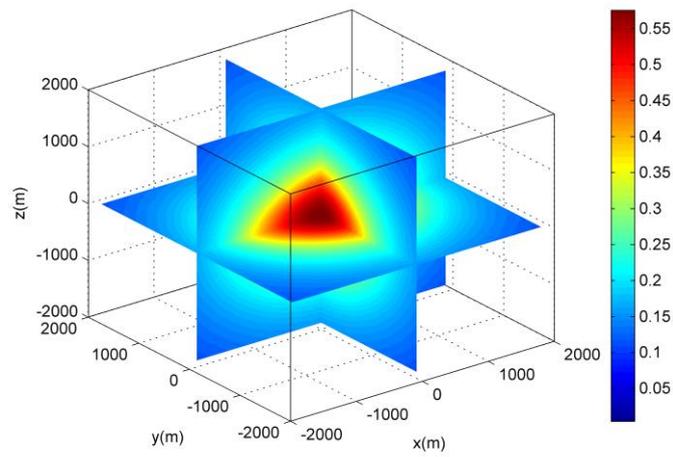

(d)

Figure 2. The structure of the gravitational potential for the primary in the coordinate system of the principal axis of inertia, the unit of the gravitational potential is m$^2$s$^{-2}$. (a) The gravitational potential in the xy plane; (b) The gravitational potential in the yz plane; (c) The gravitational potential in the zx plane; (d) the 3D structure of the gravitational potential.

To simulate the dynamical behaviors of the binary asteroidal systems with the binary octahedrons, the initial positions and moments need to set. The initial relative position and moment of the secondary are [-1.1207, 4.1826, 2.5]km and [-0.085, -0.0228, 0.0]km·s$^{-1}$, respectively. The initial angular momentums of the secondary and the primary are [-0.0015, -0.0026, 0.0048] kg·km$^2$·s$^{-1}$ and [-0.0175, -0.0386, 0.1241]kg·km$^2$·s$^{-1}$, respectively; the attitude angle of the secondary and the primary are [0.0, 0.0, 0.0]deg and [10.0, 20.0, 0.0]deg, respectively. Here all the initial relative position, initial relative velocity, initial moment, and initial angular momentum are expressed in the body-fixed frame of the primary. Thus, the initial position and the initial moment of the primary are zero. The 7/8-order Runge-Kutta method has been used to integrate the dynamical equation. Although the motion of the secondary relative to the primary is not the Keplerian motion, we still use the orbital elements to analyze the motion, and discuss the strong perturbation of the full two-body problems. The gravitational force and torque can be calculated by the expansion in series of the mutual potential of the binary polyhedral.

The gravitational force (Fahnestock and Scheeres 2006; Yu et al. 2017a, b) acting on body A and B can be expressed relative to the inertial reference frame as

$$\boldsymbol{F}_\theta^A = G \sum_{a \in A} \sum_{b \in B} \rho_a \rho_b T_a T_b \left( \frac{\partial \widehat{U}_0}{\partial \boldsymbol{A}_\theta} + \frac{\partial \widehat{U}_1}{\partial \boldsymbol{A}_\theta} + \frac{\partial \widehat{U}_2}{\partial \boldsymbol{A}_\theta} + \cdots \right) \tag{11}$$

and

$$\boldsymbol{F}_\theta^B = G \sum_{a \in A} \sum_{b \in B} \rho_a \rho_b T_a T_b \left( \frac{\partial \hat{U}_0}{\partial \boldsymbol{B}_\theta} + \frac{\partial \hat{U}_1}{\partial \boldsymbol{B}_\theta} + \frac{\partial \hat{U}_2}{\partial \boldsymbol{B}_\theta} + \cdots \right). \tag{12}$$

Here $A$ and $B$ represent the bodies and $\theta$ represent a tensor index; $\rho_a$ and $\rho_b$ are densities of the simplices $a$ and $b$, respectively; $T_a$ and $T_b$ are Jacobian determinants of the simplices $a$ and $b$, respectively; $\hat{U}_0 = \left[ \dfrac{Q}{R} \right]$, $\hat{U}_1 = \left[ -\dfrac{Q_i w^i}{R^3} \right]$, $\hat{U}_2 = \left[ -\dfrac{Q_{ij} r^{ij}}{2R^3} + \dfrac{3Q_{ij} w^i w^j}{2R^5} \right]$. $\boldsymbol{A}_\theta$ and $\boldsymbol{B}_\theta$ represent the position vectors from the origin of the inertia space to the mass centre of the bodies $A$ and $B$, respectively.

The torque acting on the each body can be expressed in its body-fixed frame by

$$\boldsymbol{M}_A = -\boldsymbol{\alpha}_P \times E^\alpha - \boldsymbol{\beta}_P \times E^\beta - \boldsymbol{\gamma}_P \times E^\gamma, \tag{13}$$

and

$$\boldsymbol{M}_B = -\boldsymbol{\alpha}_S \times E^\alpha - \boldsymbol{\beta}_S \times E^\beta - \boldsymbol{\gamma}_S \times E^\gamma. \tag{14}$$

Here the matrix $\boldsymbol{E} = \begin{bmatrix} E^\alpha & E^\beta & E^\gamma \end{bmatrix}$, and $E^\alpha$, $E^\beta$, $E^\gamma$ are column vectors of the matrix, and the element of matrix $\boldsymbol{E}$ is calculated by

$$E_{\phi\theta} = G \sum_{a \in A} \sum_{b \in B} \rho_a \rho_b T_a T_b \left( \frac{\partial \hat{U}_1}{\partial T_{\phi\theta}} + \frac{\partial \hat{U}_2}{\partial T_{\phi\theta}} + \frac{\partial \hat{U}_3}{\partial T_{\phi\theta}} + \cdots \right), \tag{15}$$

where $\boldsymbol{T} = \boldsymbol{P}^T \boldsymbol{S}$, $T_{\phi\theta}$ is the element of matrix $\boldsymbol{T}$, $\boldsymbol{P}$ is the transformation matrix from the body-fixed frame of $A$ to the inertia space, and $\boldsymbol{S}$ is the transformation matrix from the body-fixed frame of $B$ to the inertia space. The column vectors of the matrices $\boldsymbol{P}^T$ and $\boldsymbol{S}^T$ are defined by $\boldsymbol{P}^T = \begin{bmatrix} \alpha_P & \beta_P & \gamma_P \end{bmatrix}$ and $\boldsymbol{S}^T = \begin{bmatrix} \alpha_S & \beta_S & \gamma_S \end{bmatrix}$, respectively.

Let

$$\boldsymbol{v} = \left[ -P \left[ \Delta \boldsymbol{r}^{a1}, \Delta \boldsymbol{r}^{a2}, \Delta \boldsymbol{r}^{a3} \right], \; S \left[ \Delta \boldsymbol{r}^{b1}, \Delta \boldsymbol{r}^{b2}, \Delta \boldsymbol{r}^{b3} \right] \right]. \tag{16}$$

Here each of the variable in the form of $\Delta r^{(a,b)i}$ is a column vector, and is the vertex of the *i*-th face of the simplex *a* or *b*. Then in Eq. (11), using the Einstein convention of summation, we have

$$\begin{cases} \dfrac{\partial \widehat{U}_0}{\partial A_\theta} = \dfrac{Q R_\theta}{R^3} \\[6pt] \dfrac{\partial \widehat{U}_1}{\partial A_\theta} = -\dfrac{3Q_i R_\theta w^i}{R^5} - \dfrac{Q_i v_\theta^i}{R^3} \\[6pt] \dfrac{\partial \widehat{U}_2}{\partial A_\theta} = -\dfrac{3Q_{ij} r^{ij} R_\theta}{2R^5} + \dfrac{15 Q_{ij} R_\theta w^i w^j}{2R^7} - \dfrac{3Q_{ij} w^i v_\theta^j}{R^5} \\[6pt] \dfrac{\partial \widehat{U}_3}{\partial A_\theta} = \dfrac{15 Q_{ijk} r^{ij} R_\theta w^k}{2R^7} - \dfrac{3Q_{ijk} r^{ij} v_\theta^k}{2R^5} - \dfrac{35 Q_{ijk} w^i w^j w^k}{2R^9} + \dfrac{15 Q_{ijk} w^i w^j v_\theta^j}{2R^7} \end{cases} \quad (17)$$

And in Eq. (15), we have

$$\begin{cases} \dfrac{\partial \widehat{U}_1}{\partial T_{\phi\theta}} = -\dfrac{Q_i R^j D_{j\theta}^{\phi i}}{R^3} \\[6pt] \dfrac{\partial \widehat{U}_2}{\partial T_{\phi\theta}} = -\dfrac{Q_{ij} v_p^i D_{p\theta}^{\phi j}}{2R^3} + \dfrac{3Q_{ij} w^i R^p D_{p\theta}^{\phi i}}{R^5} \\[6pt] \dfrac{\partial \widehat{U}_3}{\partial T_{\phi\theta}} = \dfrac{3Q_{ijk}}{2R^5}\left(2 v_p^i D_{p\theta}^{\phi j} w^k + r^{ij} R^p D_{p\theta}^{\phi k}\right) - \dfrac{15 Q_{ijk} w^i w^j R^p D_{p\theta}^{\phi k}}{2R^7} \\[6pt] \dfrac{\partial \widehat{U}_4}{\partial T_{\phi\theta}} = \dfrac{3Q_{ijkl}}{2R^5} r^{ij} v_p^k D_{p\theta}^{\phi l} - \dfrac{15 Q_{ijkl}\left(r^{ij} w^k R^p - w^i w^j v_p^k\right) D_{p\theta}^{\phi l}}{2R^7} + \dfrac{35 Q_{ijkl} w^i w^j w^k R^p D_{p\theta}^{\phi l}}{2R^9} \end{cases} \quad (18)$$

The attitude angle means the attitude of the body-fixed frame of the octahedrons relative to the inertia frame, and is the 3-2-1 rotational order of the Euler angle. Fig. 3 shows the trajectory of the secondary relative to the primary in the inertia space. Fig. 4 gives the distance between B and A. The relative trajectory of the secondary relative to the primary is not closed, and has a significantly variety. The minimum and maximum distances between these two bodies are 2.35 km and 6.51 km, respectively. When the binary system is moving, the orbits of the two components and the attitudes of the two components relative to the inertia space all vary. If one chose the body-fixed frame of the primary as the reference frame, the motion of the secondary relative to the primary can be expressed in the translating and rotating frame. In the frame of the primary, the orbit and attitude of the secondary also have obvious changes.

To analyze the calculation precision here, we first consider the moment and energy of the binary system and their errors. Fig. 5 illustrates the moment of system relative to center of mass of A in the inertia while Fig. 6 illustrates the variety of the potential energy, kinetic energy, and total energy for the binary system. Because the system is a Hamiltonian system, the total moment and total energy are conservative. In Fig. 5, the moment of the system relative to the center of mass of A in the inertia is expressed in three different coordinate axes, and each component of the moment is a const. In Fig. 6, one can see that although the total energy is conservative, the potential energy and kinetic energy are changing during the movement of the binary system.

From Fig.7 and Fig.8, one can see that the relative error of the total angular momentum and total energy remain small. The dynamical system is integrated for $3\times 10^5$s. The relative error of the total angular momentum in inertia has the order of $10^{-10}$, and the relative error of the total energy has a smaller order of $10^{-13}$. The dynamical system consists of binary octahedrons are calculated by the two-homogeneous-polyhedral method and the 7/8-order Runge-Kutta method, thus one can conclude that the methods can be used to simulate the dynamics of the binary asteroid systems. The origin of error of the system includes two parts, one is from the truncating of the series in two-homogeneous-polyhedral method, and the other one is from the non-symplectic structure of the 7/8-order Runge-Kutta method.

Fig. 9 reveals the three-axis attitude angular velocity of the binary asteroid system in the inertia space, which includes the attitude angular velocity of the primary A and the secondary B. The components of the attitude angular velocity vary quasi-periodically. From Fig. 9(a), one can see that the influence of the attitude motion of the primary from the gravitational force of the secondary is obvious. The three-axis attitude angular velocity of the primary and the secondary vary non-periodic, which indicates that the attitude angles of the two components of the system vary non-periodic. Fig. 10 illustrates the relative momentum of B. The relative momentum of B also varies non-periodic, and its maximum amplitude is not a constant. Fig. 11 presents the angular momentum of A relative to masscenter of A in the inertia system and Fig. 12 presents the angular momentum of B relative to center of mass of B in the inertia system. Although the total angular momentum of the system is conservative, the angular momentum of A and B, as well as the components of the two bodies are changing.

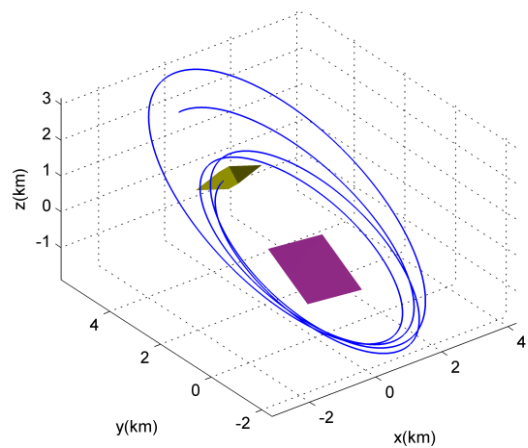

(a)

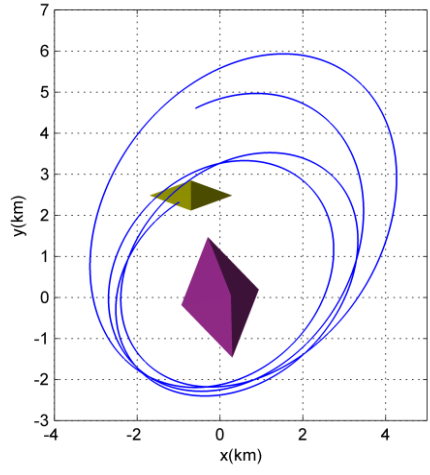

(b)

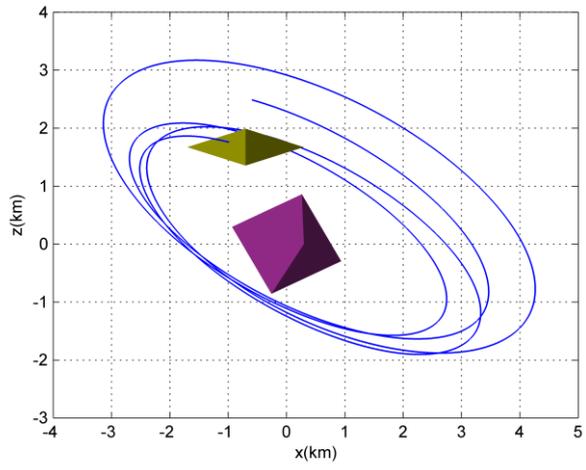

(c)

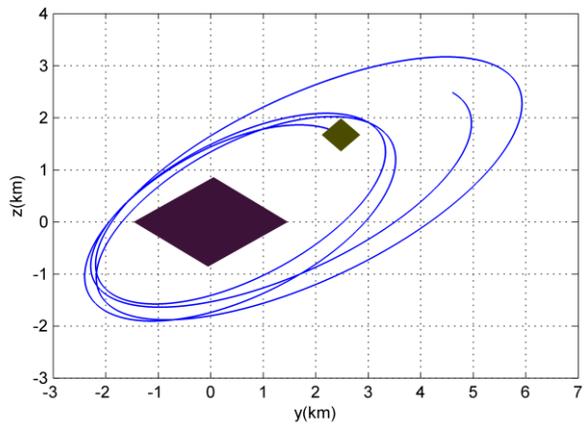

(d)

Figure 3. The trajectory of the secondary relative to the primary in the inertia space viewed from different direction. (a) the 3D view; (b) the projection viewed in the xy plane; (c) the projection viewed in the yz plane; (d) the projection viewed in the zx plane.

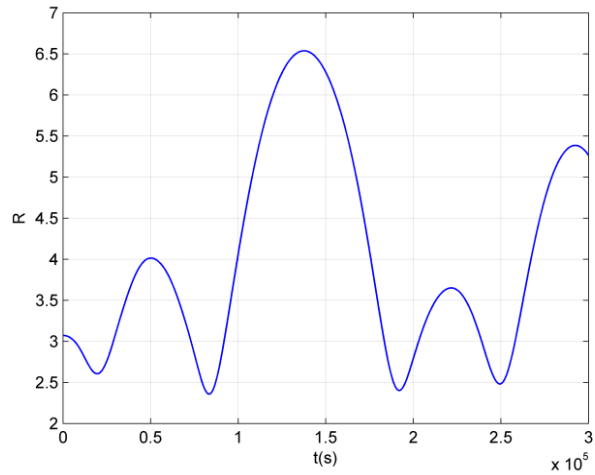

Figure 4. The distance between B and A, the unit of the distance is km.

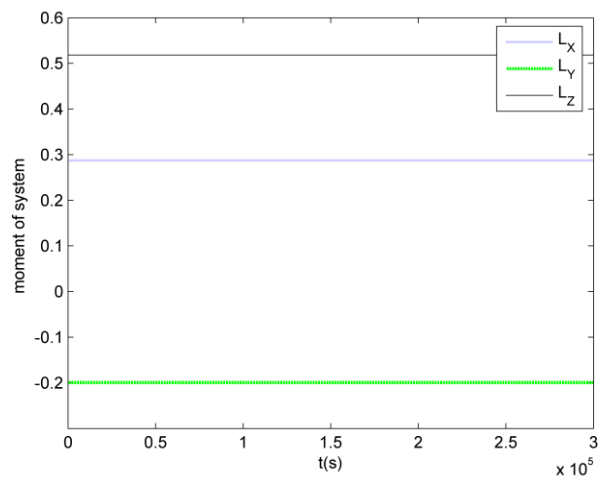

Figure 5. The moment of the system relative to the center of mass of A in the inertia, the unit of the moment is $10^{12}$ kg·m·s$^{-1}$.

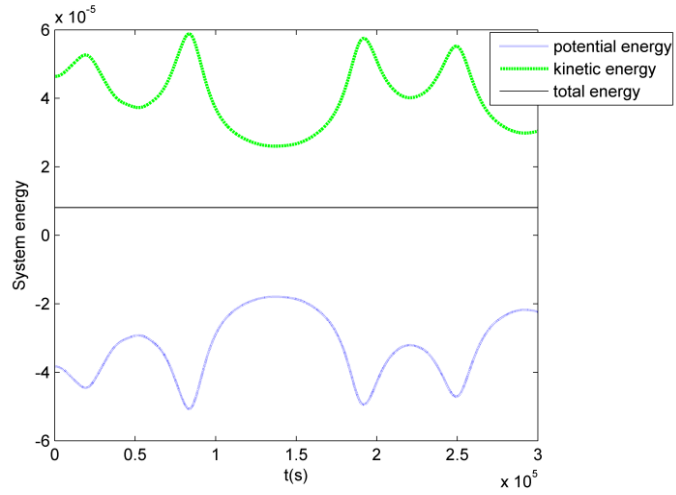

Figure 6. The variety of the potential energy, kinetic energy, and total energy for the binary system, the unit of the energy is $10^{15}$kg·m$^2$·s$^{-2}$.

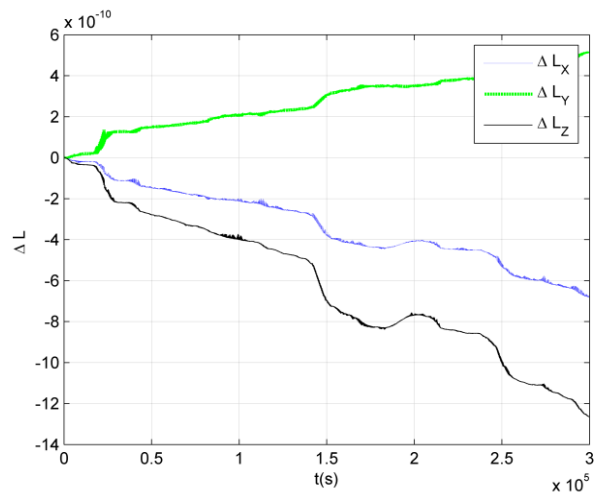

Figure 7. The relative error of the total angular momentum in the inertia space.

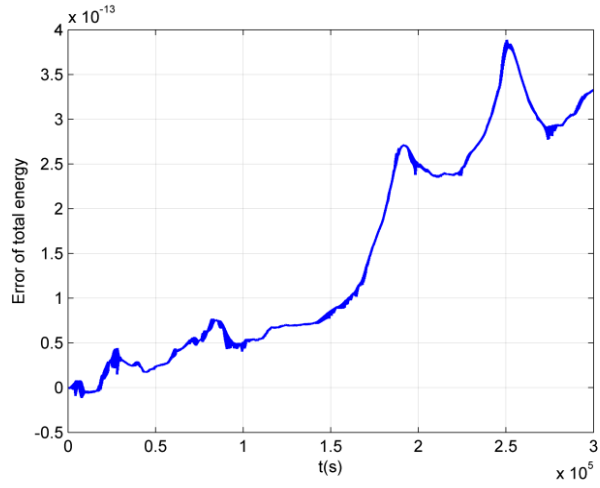

Figure 8. The relative error of the total energy for the binary system.

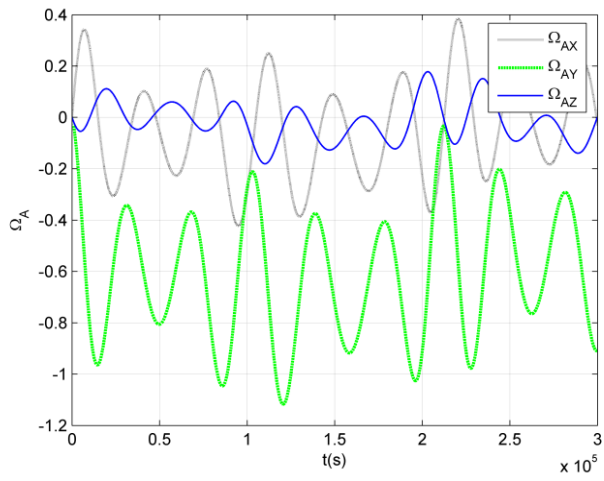

(a)

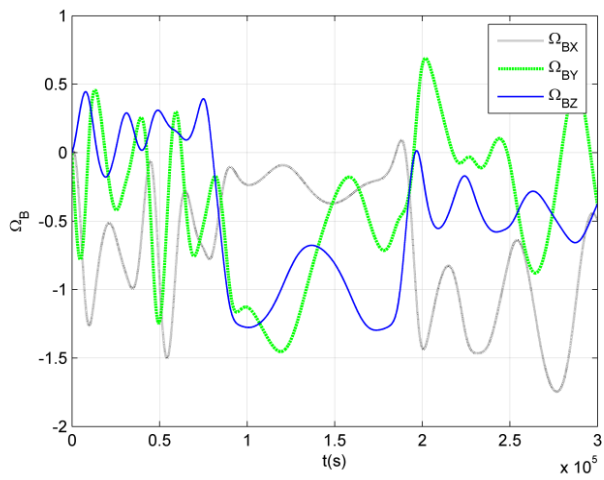

(b)

Figure 9. The three-axis attitude angular velocity of the binary system in the inertia space, the unit of the angular velocity is rad·s$^{-1}$. (a) The three-axes attitude angular velocity of the primary A; (b) The three-axes attitude angular velocity of the secondary B.

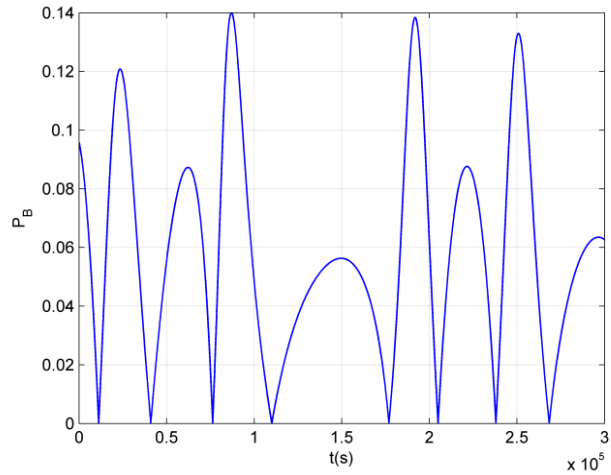

Figure 10. The relative momentum of B, the unit of the moment is $10^{12}$ kg·m·s$^{-1}$.

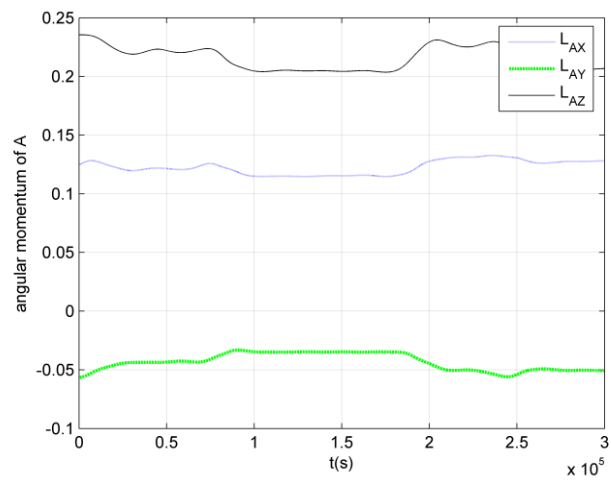

Figure 11. The angular momentum of A relative to the center of mass of A in the inertia system, the unit of the angular momentum is $10^{15}$ kg·m$^2$·rad·s$^{-1}$.

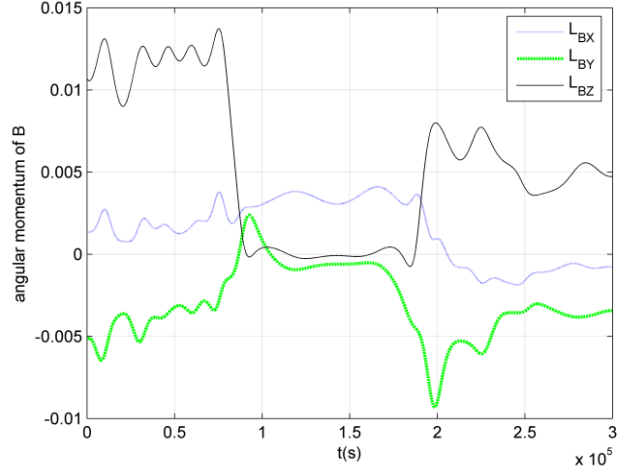

Figure 12. The angular momentum of B relative to the center of mass of B in the inertia system, the unit of the angular momentum is $10^{15}$ kg·m$^2$·rad·s$^{-1}$.

## 4 Conclusions

The dynamical behaviors of binary asteroid systems have been modeled by the binary octahedrons. The orbital parameters and attitude parameters caused by the mutual potential of the irregular shapes are calculated. We presented the binary polyhedron formulas of the mutual potential from the fourth order to the sixth order. The mutual potential of the system is calculated by the two-homogeneous-polyhedral method to the fourth order with the 7/8-order Runge-Kutta method. The result has a high precision. With the method used here, the relative error of total angular momentum in inertia has the order of $10^{-10}$ when the orbital time is $3\times10^5$s. The relative error of total energy for the binary system has the order of $10^{-13}$ when the orbital time is $3\times10^5$s.

The relative trajectory of the secondary relative to the primary is not closed. The moment of the system relative to center of mass of the primary in the inertia is conservative. The total energy of the system is also conservative. All of the orbits and attitude of the two bodies relative to the inertia space vary, including the distance, the three-axis attitude angular velocity, angular momentum, relative momentum, potential energy, and kinetic energy. The results show that the influence of the attitude motion of the primary affected by the gravitational force of the secondary is obvious. For both of the primary and the secondary, the three-axis attitude angular velocities vary non-periodic. The relative momentum of the secondary varies non-periodic. The result is helpful for understanding the complicated dynamical behaviors of the binary asteroid systems caused by the irregular shapes.

**Acknowledgements**

This project is funded by China Postdoctoral Science Foundation- General Program (No. 2017M610875), China Postdoctoral Science Foundation-Special Funding (No. 2018T110092), and the National Natural Science Foundation of China (No. 11772356).